\documentclass[prl,twocolumn,showpacs,floatfix]{revtex4-1}
\usepackage{graphicx}
\usepackage{amssymb}
\usepackage{amsmath}
\usepackage{xspace}
\usepackage{url}
\usepackage{subfigure}
\usepackage{lineno}

\let \Re \relax
\DeclareMathOperator{\Re}{Re}

\begin{document} %%%%%%%%%%%%%%%%%%%%%%%%%%%%%%%%%%%%%%%%%%%%%%%%%%%%%%%%%%
%------------------------------------------------------------------------------ 
% Title
%------------------------------------------------------------------------------
\title{Mie scattering by a charged dielectric particle}

%------------------------------------------------------------------------------ 
% Authors
%------------------------------------------------------------------------------
%------------------------------------------------------------------------------ 
% Date
%------------------------------------------------------------------------------
\author{R. L. Heinisch, F. X. Bronold, and 
H. Fehske}
\affiliation{Institut f{\"ur} Physik,
             Ernst-Moritz-Arndt-Universit{\"a}t Greifswald,
             17489 Greifswald,
             Germany}

\date{\today}
\begin{abstract}
We study for a dielectric particle the effect of surplus 
electrons on the anomalous scattering of light arising from the transverse optical phonon 
resonance in the particle's dielectric function. Excess electrons affect the polarizability of the particle 
by their phonon-limited conductivity, either in a surface layer (negative electron affinity) or the conduction band 
(positive electron affinity). We show that surplus electrons shift an extinction resonance in the infrared. This 
offers an optical way to measure the charge of the particle and to use it in a plasma as a 
minimally invasive electric probe.
\end{abstract}
\maketitle

The scattering of light by a spherical particle is a fundamental problem of electromagnetic theory. 
Solved by Mie in 1908~\cite{Mie08}, it encompasses a wealth of scattering phenomena owing to the complicated 
mathematical form of the scattering coefficients and the variety of the underlying material-specific 
dielectric constants~\cite{BH83,BW99}. While Mie scattering is routinely used as a particle size 
diagnostic~\cite{BH83}, the particle charge has not yet been determined from the Mie signal. Most particles
of interest in astronomy, astrophysics, atmospheric sciences, and laboratory experiments are however
charged
~\cite{HCL10,FR09,Mann08,Ishihara07,Mendis02}. The particle charge is a rather important parameter.
It determines the coupling of the particles among each other and to external electro-magnetic fields. An optical measurement of it 
would be extremely useful. In principle, light scattering contains information about excess electrons as their electrical conductivity modifies either the boundary condition for electromagnetic fields or the polarizability of the material \cite{KK10,KK07,BH83,BH77}. 
But how strong and in what spectral range the particle charge reveals itself by distorting the Mie signal 
of the neutral particle is an unsettled issue.

In this Letter, we revisit Mie scattering by a negatively charged dielectric particle. Where electrons are trapped on the 
particle depends on the electron affinity \(\chi\) of the dielectric, that is, the offset of the conduction band minimum to the 
potential in front of the surface. 
For \(\chi<0\), as it is the case for MgO, CaO or LiF \cite{RWK03,BKP07}, the conduction band lies above the potential outside 
the grain and electrons are trapped in the image potential induced by a surface mode associated with the transverse optical (TO) 
phonon \cite{HBF11,HBF12}. The conductivity \(\sigma_s\) of this two-dimensional electron gas is limited by the residual scattering 
with the surface mode and modifies the boundary condition 
for the electromagnetic fields at the surface of the grain. For \(\chi>0\) , as it is the case for Al$_2$O$_3$ or SiO$_2$,  electrons  
accumulate in the conduction band forming a space charge \cite{HBF12}. Its width, limited by the screening in the dielectric, 
is typically larger than a micron. For micron sized particles we can thus assume a homogeneous distribution of the excess electrons 
in the bulk. The effect on light scattering is now encoded in the bulk conductivity of the excess electrons \(\sigma_b\), which is 
limited by scattering with a longitudinal optical (LO) bulk phonon and gives rise to an additional 
polarizability per volume \(\alpha=4\pi i \sigma_b /\omega\), where \(\omega\) is the frequency of the light.
%which modifies the refractive index.
We focus on the scattering of light in the vicinity of anomalous optical 
resonances which have been identified for metal particles by Tribelsky {\it et al.} \cite{TL06,Tribelsky11}.
These resonances occur at frequencies \(\omega\) where the complex dielectric 
function \(\epsilon(\omega)=\epsilon^\prime(\omega)+i \epsilon^{\prime\prime}(\omega)\) has 
\(\epsilon^\prime<0\) and \(\epsilon^{\prime\prime}\ll1\). For a dielectric they are induced by the TO phonon 
and lie in the infrared. Using Mie theory, we show that for submicron-sized particles the extinction resonance shifts with the 
particle charge and can thus be used to determine the particle charge.

%To obtain the Mie solution for the scattering of light by a sphere the incident plane wave and the transmitted 
%and reflected waves are expanded in spherical vector harmonics. The scattering and transmission coefficients 
%connecting partial waves
%are determined by the 
%boundary conditions for the electric and magnetic fields at the surface of the particle \cite{Stratton41,BH83}. 
%For a charged particle with \(\chi<0\) the  surface charges may sustain a surface current \(\mathbf{K}\) which enters the 
%boundary condition for the magnetic field. Thus, 
%\(\hat{\mathbf{e}}_r \times \left( \mathbf{H}_i +\mathbf{H}_r -\mathbf{H}_t \right) =\frac{4\pi}{c}\mathbf{K}\), where \(i\) denotes the incident, \(r\) the reflected, and \(t\) the transmitted wave and \(c\) the speed of light\cite{BH77}.
%The surface current  \(\mathbf{K}=\sigma_s \mathbf{E}_\parallel\) is induced by the component of the electric field parallel to the surface and is proportional to the surface conductivity \(\sigma_s\) . For \(\chi>0\) the bulk surplus charge enters the refractive index \(N=\sqrt{\epsilon+\alpha}\) through its polarizability.
%Matching the fields at the boundary of a dielectric sphere with radius \(a\) 
%gives, following Bohren and Hunt \cite{BH77}, the scattering 
%coefficients 
In the framework of Mie theory, the scattering and transmission coefficients
connecting incident \((i)\), reflected \((r)\), and transmitted \((t)\) partial waves
are determined by the
boundary conditions for the electric and magnetic fields at the surface of the particle \cite{Stratton41,BH83}.
For a charged particle with \(\chi<0\) the surface charges may sustain a surface current \(\mathbf{K}\) which enters the
boundary condition for the magnetic field. Thus,
\(\hat{\mathbf{e}}_r \times \left( \mathbf{H}_i +\mathbf{H}_r -\mathbf{H}_t \right) =\frac{4\pi}{c}\mathbf{K}\),
where \(c\) is the speed of light~\cite{BH77}.
The surface current \(\mathbf{K}=\sigma_s \mathbf{E}_\parallel\) is induced by the component of the electric field
parallel to the surface and is proportional to the surface conductivity \(\sigma_s\) . For \(\chi>0\) the bulk surplus
charge enters the refractive index \(N=\sqrt{\epsilon+\alpha}\) through its polarizability.
Matching the fields at the boundary of a dielectric sphere with radius \(a\)
gives, following Bohren and Hunt \cite{BH77}, the scattering coefficients
\begin{align}
a_n^r&=\frac{\psi_n(N\rho)\psi_n^\prime (\rho)-\left[N\psi_n^\prime(N\rho)-i\tau \psi_n(N\rho)\right] 
\psi_n(\rho) }{\left[N\psi_n^\prime(N\rho)-i\tau \psi_n(N\rho)\right]\xi_n(\rho)-\psi_n(N\rho)\xi_n^\prime(\rho)}~, \\
b_n^r&=\frac{\psi_n^\prime(N\rho)\psi_n(\rho)-\left[N\psi_n(N \rho)+i\tau \psi_n^\prime(N\rho)\right]\psi_n^\prime(\rho)}{\left[N\psi_n(N\rho)+i\tau \psi_n^\prime(N\rho) \right]\xi_n^\prime(\rho)-\psi_n^\prime(N\rho)\xi_n(\rho)}
\label{MieEq}
\nonumber
\end{align}
where for \(\chi<0\) (\(\chi>0\)) the dimensionless surface conductivity \(\tau(\omega)=4\pi\sigma_s(\omega)/c\)
(\(\tau=0\)) and the refractive index \(N=\sqrt{\epsilon}\) (\(N=\sqrt{\epsilon+\alpha}\)), the 
size parameter \(\rho= ka=2\pi a /\lambda\), where \(k\) is the wavenumber,  
\(\psi_n(\rho)=\sqrt{\pi \rho/2} J_{n+1/2}(\rho)\), \(\xi_n(\rho)=\sqrt{\pi \rho/2} H^{(1)}_{n+1/2}(\rho)\)
with \(J_n(\rho)\) the Bessel and \(H_n^{(1)}(\rho)\) the Hankel function of the first kind. 
As for uncharged particles the   extinction efficiency becomes \(Q_t=-(2/\rho^2)\sum_{n=1}^\infty \left(2n+1\right) \Re \left(a_n^r+b_n^r \right)\).
Any effect of the surplus electrons on the scattering of light, encoded in \(a_n^r\) and \(b_n^r\), is due to the 
surface conductivity (\(\chi<0\)) or the bulk conductivity (\(\chi>0\)).

For \(\chi<0\) we describe the surface electron film in 
a planar model to be justified below. For the dielectrics we consider, the low-frequency dielectric function is 
dominated by an optically active TO phonon with frequency \(\omega_{TO}\). For the modelling of the surface electrons 
it suffices to approximated it by 
\(\epsilon(\omega)=1+\omega_{TO}^2(\epsilon_0-1)/(\omega_{TO}^2-\omega^2)\), 
where \(\epsilon_0\) is the static dielectric constant,
and allows for a TO 
surface mode whose frequency is given by \(\epsilon(\omega_s)=-1\) leading to 
\(\omega_s=\omega_{TO}\sqrt{(1+\epsilon_0)/2}\) \cite{EM73}.
The coupling of the electron to this surface mode 
consists of a static and a dynamic part \cite{Barton81}.  The former 
leads to the image potential \(V=-\Lambda_0 e^2 /z\) with \(\Lambda_0=(\epsilon_0-1)/(4(\epsilon_0+1))\) 
supporting a series of bound Rydberg states whose wave functions read 
%\begin{align}
%\phi_{n \mathbf{k}}(\mathbf{x},z)=\frac{1}{\sqrt{A}}e^{i\mathbf{k}\mathbf{x}}\sqrt{\frac{\Lambda_0}{a_B n n!^2}} 
%W_{n,1/2}\left(\frac{2\Lambda_0 z}{n a_B} \right) \label{wavefunction}
%\end{align}
$\phi_{n \mathbf{k}}(\mathbf{x},z)=\sqrt{\Lambda_0/Aa_B n n!^2}e^{i\mathbf{k}\mathbf{x}} 
W_{n,1/2}\left(2\Lambda_0 z/n a_B\right)$
with \(a_B\) the Bohr radius, \(\mathbf{k}=(k_x,k_y)\), \(\mathbf{x}=(x,y)\), and \(A\) the surface area.
% which
% Transitions between image states 
%- allowing trapping and release of electrons - are due to a longitudinal acoustic bulk phonon responsible for 
%surface vibrations \cite{HBF11}. 
Since trapped electrons are thermalized with the surface and the spacing between
Rydberg states is large compared to $k_BT$, they occupy only the lowest image band \(n=1\). Assuming 
a planar surface is justified provided the de Broglie wavelength \(\lambda_{dB}\) of the electron on the surface 
is smaller than the radius \(a\) of the sphere. For a surface electron with energy \(E_{kin}/k_B=300 K \) one 
finds \(\lambda_{dB}\approx 10^{-8} cm\). Thus, for particle radii \(a>10 nm\) the plane-surface 
approximation is justified. The dynamic interaction enables momentum relaxation parallel to the surface 
and limits the surface conductivity. 
Introducing 
annihilation operators \(c_\mathbf{k}\) and \(a_\mathbf{Q}\) for electrons and phonons, the Hamiltonian
describing the dynamic electron-phonon coupling in the lowest image band reads, 
\(H=\sum_\mathbf{k} \epsilon_\mathbf{k} c_\mathbf{k}^\dagger c_\mathbf{k} + \hbar \omega_s \sum_\mathbf{Q} 
a_\mathbf{Q}^\dagger a_\mathbf{Q}\nonumber +H_{int}\) \cite{KI95} with
%\begin{align}
%H_{int}=\frac{1}{\sqrt{A}}\sum_{\mathbf{k},\mathbf{Q}}M_{\mathbf{k},\mathbf{Q}} c^\dagger_{\mathbf{k}+\mathbf{Q}}\left(a_\mathbf{Q}-a^\dagger_{-\mathbf{Q}} \right)c_\mathbf{k} \text{ ,}
%\end{align}
$H_{int}=\sum_{\mathbf{k},\mathbf{Q}}(M_{\mathbf{k},\mathbf{Q}}/\sqrt{A}) 
c^\dagger_{\mathbf{k}+\mathbf{Q}}(a_\mathbf{Q}-a^\dagger_{-\mathbf{Q}})c_\mathbf{k}$,
where the matrix element is given by ($m$ is the electron mass)
\begin{align}
M_{\mathbf{k} \mathbf{Q}}= \frac{2e\sqrt{\pi \Lambda_0 \hbar^3}}{m \sqrt{\omega_s Q}} 
\left( \frac{2 \Lambda_0}{Q a_B +2\Lambda_0} \right)^3  \left[\mathbf{Q}\mathbf{k}+\frac{Q^2}{2} \right] ~.
\end{align}
Within the memory function approach \cite{GW72} the surface conductivity can be written as 
\begin{align}
\sigma_s(\omega)=\frac{e^2 n_s}{m} \frac{i}{\omega +M(\omega)} \label{surfcon}
\end{align}
%$\sigma_s(\omega)=\frac{e^2 n_s}{m} \frac{i}{\omega +M(\omega)}$
with \(n_s\) the surface electron density. Up to second order in the electron-phonon coupling 
the memory function  
\begin{align}
M(\omega)&=\frac{\sqrt{m \omega_s \delta} e^2 \Lambda_0}{\sqrt{2\pi \hbar^3}  } 
\int_{-\infty}^\infty \mathrm{d}\bar{\nu} \frac{j(-\bar{\nu}) -j(\bar{\nu})}{\bar{\nu} (\bar{\nu}-\nu-i0^+)}~,  \label{surfmem}
\\
%\end{align}
%with
%\begin{align}
j(\nu)&=\frac{e^\delta}{e^\delta-1}  |\nu+1|^3 e^{-\delta(\nu+1)/2} I_\frac{\gamma}{\sqrt{|\nu+1|}}\left(\frac{\delta|\nu+1|}{4}\right) \nonumber \\
+&\frac{1}{e^\delta-1}|\nu-1|^3 e^{-\delta(\nu-1)/2} I_\frac{\gamma}{\sqrt{|\nu-1|}}\left(\frac{\delta|\nu-1|}{4}\right)~,
%\nonumber
\end{align}
where \(\nu=\omega/\omega_s\), \(\delta=\beta\hbar\omega_s\), \(\gamma=\sqrt{2 \Lambda_0^2\hbar/a_B^2 m \omega_s}\),
and \(I_a(x)=\int_0^\infty \mathrm{d}t  e^{-x\left(1/t+t\right)} a^6/(a+\sqrt{t})^6  \)
which for low temperature, that is \(x\rightarrow \infty\), has the asymptotic form
\(I_a(x)\sim \sqrt{\pi/x} e^{-2x} a^6/(1+a)^6  \text{.} \)
Since \(M(\omega)\) is independent of \(n_s\) the surface conductivity is proportional to \(n_s\).
\begin{figure}[t]
\includegraphics[width=\linewidth]{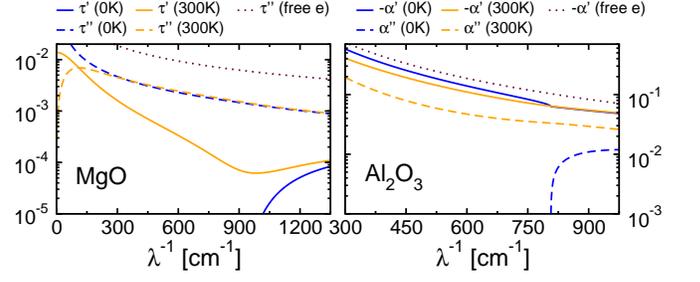}
\caption{Dimensionless surface conductivity \(\tau=\tau^{\prime}+i\tau^{\prime \prime}\) for MgO
for \(n_s=10^{13}cm^{-2}\)  (left) and polarizability of excess electrons \(\alpha=\alpha^\prime+i\alpha^{\prime\prime}\) for Al$_2$O$_3$
for \(n_b=3\times 10^{17}cm^{-3}\) (right) as a function of the inverse wavelength \(\lambda^{-1}\).}
\label{fig_par}
\end{figure}

For \(\chi>0\) the bulk conductivity is limited by a longitudinal optical (LO) phonon with frequency \(\omega_{LO}\). The coupling of the electron to this mode is described by \(H_{int}=\sum_{\mathbf{k},\mathbf{q}} M c_{\mathbf{k}+\mathbf{q}}^\dagger c_\mathbf{k} \left(a_\mathbf{q}+a_{-\mathbf{q}}^\dagger \right)/\sqrt{V}q\) \cite{Mahan90}, 
where \(M=\sqrt{2\pi e^2\hbar\omega_{LO}\left(\epsilon_\infty^{-1}-\epsilon_0^{-1} \right)}\).  Employing the memory function approach, 
the bulk conductivity is given by  Eq. (\ref{surfcon}) where \(n_s\) is replaced by the bulk electron density \(n_b\) and \(m\) by the 
conduction band effective mass \(m^\ast\), the prefactor of the memory function (Eq. (\ref{surfmem})) is then 
\(4 e^2 \sqrt{m^\ast \omega_{LO} \delta}(\epsilon_\infty^{-1}-\epsilon_0^{-1})/(3 \sqrt{(2\pi \hbar)^3})\), and \begin{align}
j(\nu)=&\frac{e^\delta}{e^\delta-1}  |\nu+1| e^{-\delta(\nu+1)/2}K_1\left(\delta|\nu+1| /2 \right) \nonumber \\
+&\frac{1}{e^\delta-1}|\nu-1| e^{-\delta(\nu-1)/2}K_1\left(\delta|\nu-1|/2 \right) \text{ ,}
\end{align}
where \(\nu=\omega/\omega_{LO}\), \(\delta=\beta\hbar\omega_{LO}\), and \(K_1(x)\) is a modified Bessel function. For low temperature, i.e. \(\delta \rightarrow \infty\) \(j(\nu)\sim \sqrt{\pi/\delta} \sqrt{|\nu+1|} \theta(-\nu-1)\).

\begin{figure}[t]
\includegraphics[width=\linewidth]{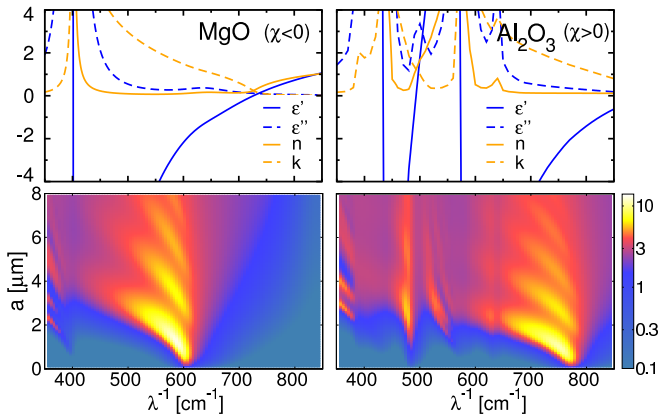}
\caption{Dielectric constant \(\epsilon=\epsilon^\prime+ i \epsilon^{\prime \prime}\),
refractive index \(N=n+ik\) (top) and extinction efficiency \(Q_t\) (bottom) depending on the particle radius \(a\) for MgO and Al$_2$O$_3$ as 
a function of the inverse wavelength
\(\lambda^{-1}\).}
\label{fig_eps}
\end{figure}

To exemplify light scattering by a charged dielectric particle we will consider a MgO (\(\chi<0\)) and an Al$_2$O$_3$ (\(\chi>0\)) 
particle \cite{matpar}.
The charge effect on scattering is controlled by the dimensionless surface 
conductivity \(\tau=\tau^\prime+i\tau^{\prime \prime}\) (for \(\chi<0\)) or the excess electron 
polarizability \(\alpha=\alpha^\prime+i\alpha^{\prime \prime}\) (for \(\chi>0\)), both shown in Fig. \ref{fig_par}, which 
are small even for a highly charged particle with \(n_s=10^{13}cm^{-2}\) (corresponding to \(n_b=3\times 10^{17}cm^{-3}\) 
for \(\chi>0\) and \(a=1\mu m\)).  The electron-phonon coupling reduces \(\tau^{\prime \prime}\) and \(\alpha^\prime\) compared 
to a free electron gas where \(M(\omega)=0\), implying \(\tau^\prime=0\) and \(\alpha^{\prime \prime}=0\). 
For \(T=0K\), \(\tau^\prime=0\) (\(\alpha^{\prime \prime}=0\)) for \(\lambda^{-1}<\lambda_s^{-1}=909cm^{-1}\), the inverse 
wavelength of the surface phonon ( \(\lambda^{-1}<\lambda_{LO}^{-1}=807cm^{-1}\), the inverse wavelength of the bulk LO phonon) 
since light absorption is only possible above the surface (bulk LO) phonon frequency. 
At room temperature \(\tau^{\prime \prime}\) and \(\alpha^\prime\) still outweigh \(\tau^\prime\) and \(\alpha^{\prime\prime}\). 
The temperature effect on  \(\tau^{\prime \prime}\) is less apparent for \(\lambda^{-1}>300cm^{-1}\) than for \(\alpha^\prime\) 
but for  \(\lambda^{-1}<300cm^{-1}\) a higher temperature lowers  \(\tau^{\prime \prime}\) considerably.
The upper panel of Fig. \ref{fig_eps} shows the complex dielectric constant \(\epsilon\) and the refractive index \(N\). For 
MgO we use a two-oscillator fit for \(\epsilon\) \cite{JKP66,Palik85}.
In the infrared, \(\epsilon\) is dominated by a TO phonon at 
\(401cm^{-1}\). The second phonon at \(640cm^{-1}\) is much weaker, justifying our model for the image potential 
based on one dominant phonon. 
Far above the highest TO phonon, that is, for \(\lambda^{-1} > 800 cm^{-1}\) (\(\lambda^{-1} > 900 cm^{-1}\)) for MgO 
(Al$_2$O$_3$) \(\epsilon^\prime >0\) and \(\epsilon^{\prime \prime} \ll 1\). For these wavenumbers a micron sized grain would 
give rise to a typical Mie plot exhibiting 
interference and ripples due to the functional form of \(a_n^r\) and \(b_n^r\) and not 
due to the dielectric constant.
Surplus electrons would not alter the extinction in this region because \(|\epsilon| \gg |\tau|\) and \(|\epsilon|\gg|\alpha| \). 

\begin{figure}
\begin{minipage}{0.49\linewidth}
\rotatebox{270}{\includegraphics[width=1.2\linewidth]{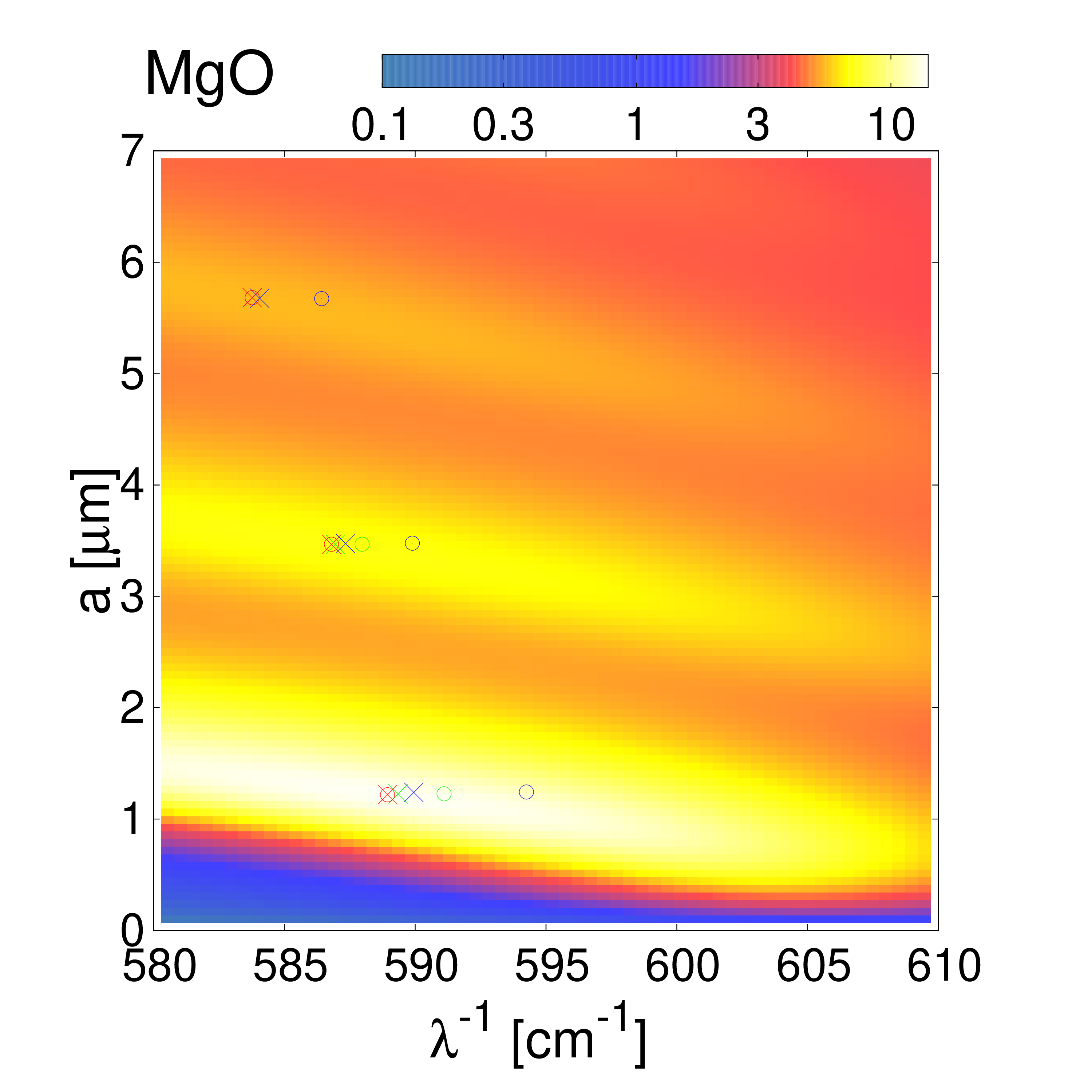}}
\end{minipage}
\begin{minipage}{0.49\linewidth}
\rotatebox{270}{\includegraphics[width=1.2\linewidth]{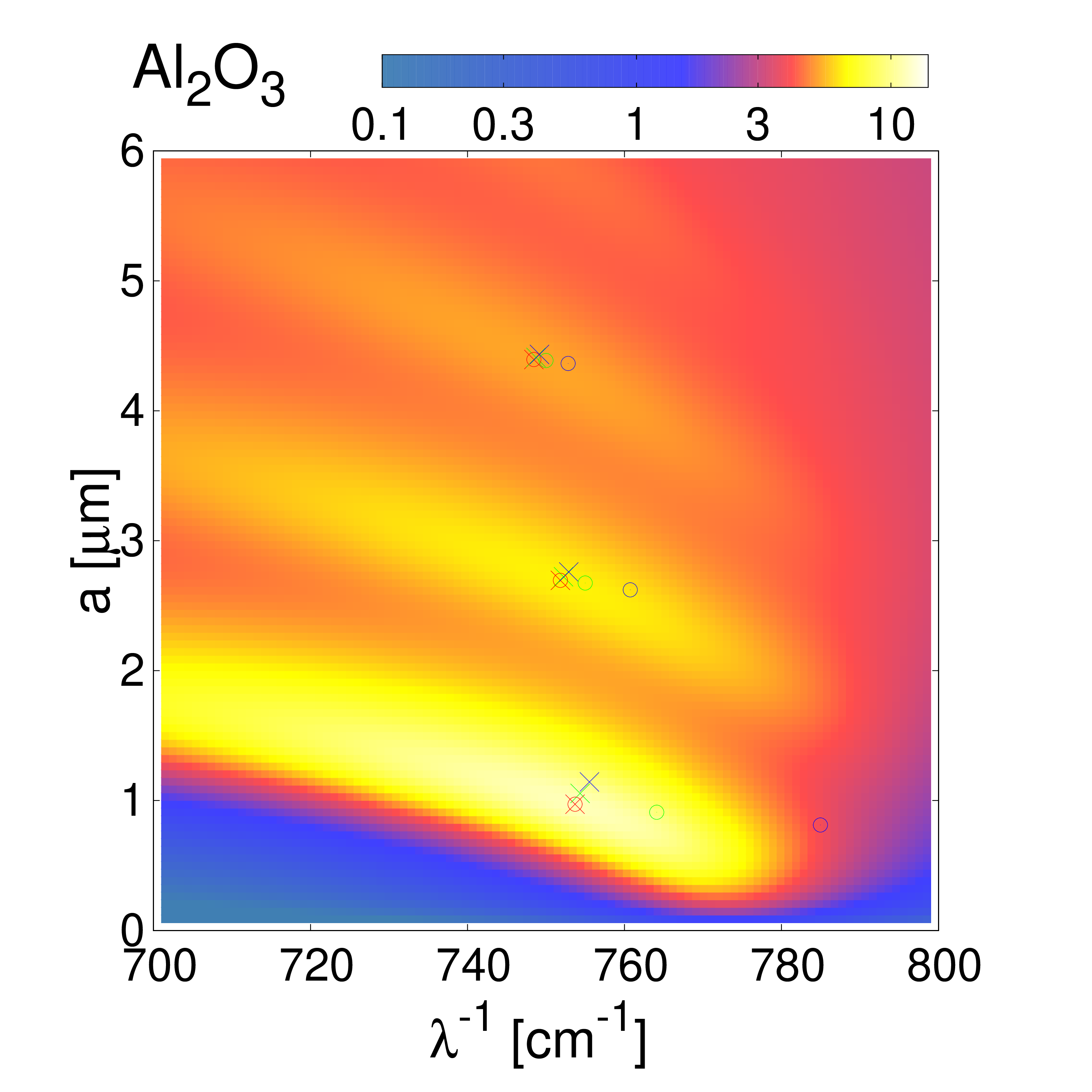}}
\end{minipage}
\caption{Magnification of the extinction resonances depending on \(\lambda^{-1}\) and \(a\). Crosses indicate their maxima for
\(n_s=0\) (red), \(2\times 10^{13}\) (green), and  \(5\times 10^{13}cm^{-2}\)
(blue) at \(T=300K\) for MgO (left panel) and for \(n_b=3n_s/a\) for Al$_2$O$_3$ (right panel). Open circles indicate the maxima for free electrons.}
\label{fig_detail}
\end{figure}

To observe a stronger dependence of extinction on the parameters \(\epsilon\) and 
\(\tau\) or \(\alpha\), we turn to  \(400cm^{-1} < \lambda^{-1} < 700cm^{-1}\)  for MgO (\(700cm^{-1} < \lambda^{-1} < 900cm^{-1}\) for Al$_2$O$_3$) where \(\epsilon^\prime<0\) and 
\(\epsilon^{\prime\prime}\ll1\) allowing for optical resonances, sensitive to even small changes 
in \(\epsilon\). They correspond to resonant excitation of transverse surface modes of the sphere \cite{FK68}. For a metal particle the resonances are due to plasmons and lie in the ultraviolet \cite{TL06,Tribelsky11}. For a dielectric the TO phonon induces them. As the polarizability of excess electrons, encoded in \(\tau\) or \(\alpha\), is larger at low frequency, the resonances of a dielectric particle, lying in the infrared, should be more susceptible to surface charges.  The lower panel of Fig. \ref{fig_eps} shows a clearly distinguishable series of resonances in the extinction efficiency.
The effect of negative excess charges is shown by the crosses in Fig. \ref{fig_detail}. The extinction maxima are shifted to higher \(\lambda^{-1}\) 
for both surface and bulk excess electrons. For comparison the circles show the shift for a free electron gas. 
%Clearly, the electron-phonon coupling 
%limits the shift. 
The effect is strongest 
for the first resonance, where a surface electron density \(10^{13} cm^{-2}\) (or an equivalent bulk charge), realized for instance  
in dusty plasmas \cite{FGP11}, yields a shift of a few wavenumbers.

\begin{figure}
\includegraphics[width=1.0\linewidth]{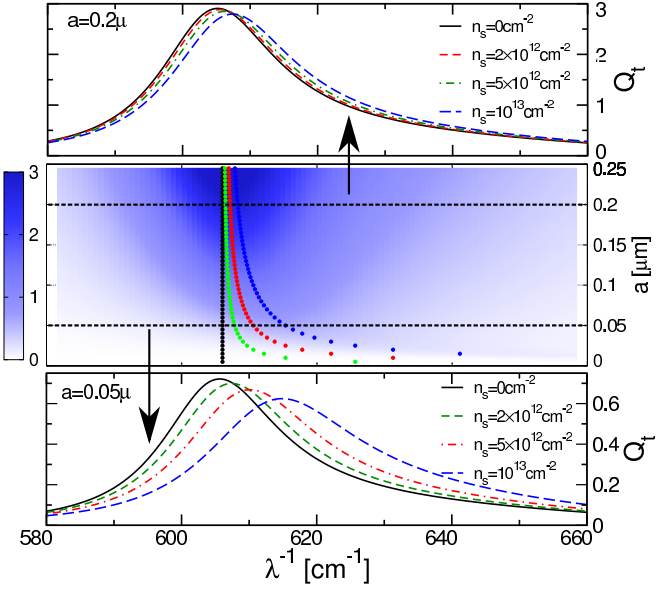}
\caption{Middle panel: Extinction efficiency \(Q_t\) as a function of the inverse wavelength \(\lambda^{-1}\) and
the radius \(a\) for a MgO particle with \(n_s= 10^{13} cm^{-2}\) and \(T=300 K\). The
dotted lines indicate the extinction maximum for \(n_s=0\) (black),
\(2\times 10^{12}\) (green), \(5\times 10^{12}\) (red), and \(10^{13} cm^{-2}\) (blue)
obtained from (\ref{resloc}). Top and bottom panel: Extinction efficiency \(Q_t\) as a function of \(\lambda^{-1}\) for \(a=0.2\mu m\) (top) and \(a=0.05\mu m\) (bottom) for different surface electron
densities. }
\label{fig_tail}
\end{figure}

The shift can be more clearly seen in Fig. \ref{fig_tail} where the tail of the first 
resonance is plotted for MgO on an enlarged scale. The main panel 
shows the extinction efficiency for \(n_s= 10^{13} cm^{-2}\) with its maxima indicated by blue dots. 
Without surface charge the resonance is at \(\lambda^{-1}=606cm^{-1}\) for \(a<0.25\mu m\). 
For a charged particle it moves to higher \(\lambda^{-1}\) and this effect becomes stronger the smaller 
the particle is. The line shape of the extinction resonance for fixed particle size is 
shown in the top and bottom panels for \(a=0.2\mu m\) and \(a=0.05\mu m\), respectively. For comparison 
data for other surface charge densities are also shown. 
%Clearly, surface electrons induce
%a shift of the extinction resonance. 
Figure \ref{fig_tail} also suggests that the resonance shift is even more significant 
for particles with radius \(a<0.01\mu m\) where the planar model for the image states is inapplicable. An extension of our model, 
guided by the study of multielectron bubbles in helium \cite{TSD07}, requires surface phonons, image potential and electron-phonon 
coupling for a sphere. Due to its insensitivity to the location of the excess electrons, we expect qualitatively the same resonance 
shift for very small grains.

\begin{figure}[t]
\includegraphics[width=\linewidth]{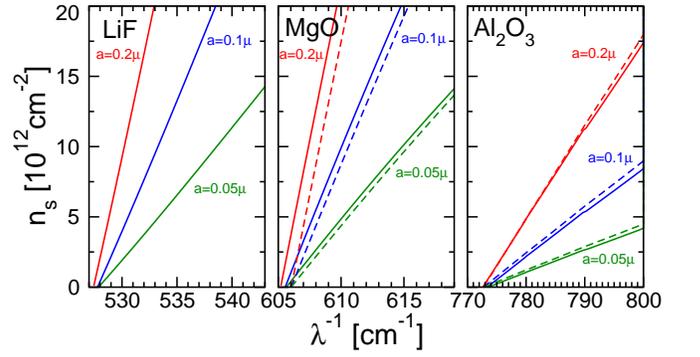}
\caption{Position of the extinction resonance depending on the surface charge \(n_s\) for LiF
, MgO and Al$_2$O$_3$ (for equivalent bulk charge \(n_b=3 n_s/a\)) particles
with different radii \(a\). Solid (dashed) lines are obtained from the Mie contour (Eq. (\ref{resloc})).}
\label{fig_triple}
\end{figure}

As we are considering particles small compared to $\lambda$ we expand the scattering coefficients for 
small \(\rho\). To ensure that in the limit of an uncharged grain, that is, for \(\tau \rightarrow 0\), \(a_n^r\) 
and \(b_n^r\) converge to their small \(\rho\) expansions \cite{Stratton41}, we substitute \(t=\tau/\rho\) 
before expanding the coefficients. Up to \(\mathcal{O}(\rho^3)\) this yields \(a_1^r=a_2^r=b_2^r=0\) 
and only \(b_1^r\sim \mathcal{O}(\rho^3)\) contributes. Then the extinction efficiency,
\begin{align}
Q_t= \frac{12\rho\left(\epsilon^{\prime\prime}+\alpha^{\prime \prime}+2\tau^{\prime }/\rho \right)}
{\left(\epsilon^\prime+\alpha^\prime+2- 2\tau^{\prime\prime}/\rho \right)^2+
\left(\epsilon^{\prime \prime}+\alpha^{\prime \prime}+2 \tau^\prime/\rho \right)^2}~,
\end{align}
where the excess charges enter either through \(\tau\) with \(\alpha=0\) for \(\chi<0\) or through \(\alpha\) 
with \(\tau=0\) for \(\chi>0\).
%Excess charges enter either by \(\tau\) (\(\chi<0\)) or \(\alpha\) (\(\chi>0\)). 
For \(\tau, \alpha \rightarrow 0\) this gives the limit of Rayleigh 
scattering. The resonance is located at the wavenumber where
\begin{align}
\epsilon^\prime+\alpha^\prime +2- 2\tau^{\prime \prime}/\rho =0 \text{ } \label{resloc}
\end{align}
and has a Lorentzian shape, already apparent from Fig. \ref{fig_tail},
 provided \(\epsilon^{\prime \prime}\) and \(\tau^\prime\) (or \(\alpha^{\prime\prime}\)) vary only 
 negligibly near the resonance wavelength.
For an uncharged surface the resonance is at \(\lambda_0^{-1}\) for which \(\epsilon^\prime =-2\). For \(\chi<0\) the shift of the resonance is proportional to \(\tau^{\prime\prime}\) and thus to \(n_s\),
 provided \(\epsilon^\prime\) is well approximated linearly in \(\lambda^{-1}\) and \(\tau^{\prime \prime}\) does not vary 
significantly near \(\lambda_0^{-1}\). In this case, we substitute in (\ref{resloc}) the
expansions \(\epsilon^\prime=-2+c_\epsilon(\lambda^{-1}-\lambda_0^{-1})\) and \(\tau^{\prime \prime}=c_\tau n_s\) 
where \(c_\epsilon=\frac{\partial \epsilon^\prime}{\partial \lambda^{-1}}|_{\lambda_0^{-1}}\) and 
\(c_\tau=\frac{\tau^{\prime \prime}}{n_s}|_{\lambda_0^{-1}}\). Then the resonance is located at 
\(\lambda^{-1}=\lambda_0^{-1}+c_\tau n_s/(\pi c_\epsilon a \lambda_0^{-1})\). For \(\chi>0\) the resonance is located at \(\lambda^{-1}=\lambda_0^{-1}-c_\alpha n_b/c_\epsilon \) where \(c_\alpha=\frac{\alpha^{\prime}}{n_b}|_{\lambda_0^{-1}}\).  The dotted lines in 
Fig. \ref{fig_tail} give the location of the resonance obtained from Eq. (\ref{resloc}) for MgO, where \(\lambda_0^{-1}=606cm^{-1}\) 
for several surface 
electron densities. They agree well with the underlying contour calculated from the exact Mie solution,
as exemplified for \(n_s= 10^{13}cm^{-2}\). 
 The proportionality of the resonance shift to \(n_s\) (\(n_b\)) can also be seen in
Fig. \ref{fig_triple} where we plot on the abscissa the shift of the extinction resonance arising 
from the surface electron density given on the ordinate for LiF \cite{matpar},  MgO (\(\chi<0\)) and Al$_2$O$_3$ (\(\chi>0\)).
Both bulk and surface electrons lead to a resonance shift. To illustrate the similarity of the shift we consider 
%the resonance condition 
(\ref{resloc}) for free electrons, which then becomes \(\epsilon^{\prime}-2N_ee^2/(m a^3\omega^2)=-2\) for \(\chi<0\) and 
\(\epsilon^\prime-3N_ee^2/(m^\ast a^3\omega^2)=-2\) for \(\chi>0\); \(N_e\) is the number of excess electrons. The effect 
of surface electrons is weaker by a factor \(2 m^\ast /3 m\) where the \(2/3\) arises from geometry as only the parallel 
component of the electric field acts on the spherically confined electrons. Most important, however, \(\tau/\rho\) and \(\alpha\) 
enter the resonance condition on the same footing showing that the shift is essentially an electron density effect on 
the polarizability of the grain. We therefore expect the shift to prevail also for electron distributions between the two limiting cases 
of a surface and a homogeneous bulk charge. 

To conclude, our results suggest that for dielectric particles showing anomalous optical resonances the 
extinction maximum in the infrared can be used to determine the particle charge (see 
Fig.~\ref{fig_triple}). For dusty plasmas this can be rather attractive because established 
methods for measuring the particle charge~\cite{CJG11,KRZ05,TLA00} require plasma parameters 
which are not precisely known whereas the charge measurement by Mie scattering does not.
Particles with surface (negative electron affinity \(\chi\), e.g. MgO, LiF) as well as bulk excess 
electrons (\(\chi>0\) e.g. Al$_2$O$_3$) show the effect and
could serve as model systems for 
sub-micron sized dust in space, the laboratory, and the atmosphere. These 
particles could be also used as minimally invasive electric probes in a plasma, which collect electrons 
depending on the local plasma environment. Determining their charge from Mie scattering and the forces acting on them 
by conventional means~\cite{CJG11,KRZ05,TLA00} would provide a way to extract plasma parameters locally.

We acknowledge support by the Deutsche Forschungsgemeinschaft through 
SFB-TRR 24, Project B10.

%\bibliography{ref} 
%\bibliographystyle{apsrev}

\end{document}